\documentclass[
  preprintnumbers, floatfix, letterpaper, aps, prd,
  epsfig, nofootinbib, twocolumn
]{revtex4-1}

\usepackage{bm, graphicx, dcolumn, epstopdf, epsf, latexsym,
  mathbbol, amssymb, amsmath, slashed, mathrsfs, mathcomp, simplewick}
\usepackage[center]{subfigure}
\usepackage{multirow}
\usepackage{makecell}
\usepackage{booktabs}
\usepackage{placeins}

\usepackage{xcolor}
\usepackage{soul}
\sethlcolor{yellow}

\usepackage[colorlinks, linkcolor=blue, citecolor=blue, urlcolor=blue]{hyperref}
\usepackage{orcidlink}
\soulregister\cite7

\newcommand{\bq}{\begin{equation}}
\newcommand{\eq}{\end{equation}}
\newcommand{\bqn}{\begin{eqnarray}}
\newcommand{\eqn}{\end{eqnarray}}

\begin{document}
\pagenumbering{arabic}
\allowdisplaybreaks

\title{Thin Accretion Disk around a Schwarzschild Black Hole immersed in a dark matter halo}

\author{Dengyu Zhao${}^{a, b}$\,\orcidlink{0009-0006-1637-9379}}
\email{211124090033@zjut.edu.cn}

\author{Hamza Rehman${}^{a, b}$\,\orcidlink{0009-0004-1553-9022}}
\email{hamzarehman244@zjut.edu.cn, hamzarehman244@gmail.com}

\author{Tao Zhu${}^{a, b}$\,\orcidlink{0000-0003-2286-9009}}
\email{zhut05@zjut.edu.cn; Corresponding author}

\affiliation{${}^{a}$ Institute for Theoretical Physics and Cosmology, Zhejiang University of Technology, Hangzhou 310023, China}
\affiliation{${}^{b}$ United Center for Gravitational Wave Physics (UCGWP), Zhejiang University of Technology, Hangzhou, 310023, China}

\date{\today}

  \begin{abstract}
  In realistic astrophysical environments, black holes are generally not isolated but may be immersed in dark matter halos. Therefore, investigating the effects of dark matter on black hole spacetimes and their surrounding accretion disks is important. We adopted the Einstein cluster model proposed by Cardoso et al, and incorporated the Jaffe profile to  describe the effects of dark matter on the spacetime geometry of a static and spherically symmetric black hole. We examined the radial epicyclic and Keplerian orbital frequencies and calculated the energy flux, temperature, and emission spectrum of the thin accretion disk using the Novikov–Thorne model. Increasing the halo mass reduces both frequencies, whereas increasing the halo scale radius brings them closer to their Schwarzschild values. Finally, we found that the influence of the dark matter halo on the accretion disk is negligible for the adopted astrophysically relevant parameter scales. However, decreasing the halo scale radius or increasing its mass leads to more pronounced modifications of the accretion disk properties.
  \end{abstract}

\maketitle

\section{Introduction}
\renewcommand{\theequation}{1.\arabic{equation}} \setcounter{equation}{0}
Black holes are extremely compact objects in the universe and ideal laboratories for testing general relativity. Due to their strong gravity, black holes pull in surrounding matter. Matter with sufficient angular momentum does not fall directly into the black hole but instead orbits around it, forming an accretion disk \cite{Shakura:1973boa}. Black holes can also power jets through the Blandford-Znajek mechanism \cite{Blandford:1977ds}. In 2019, the Event Horizon Telescope (EHT) Collaboration released the first image of the supermassive black hole in the center of the M87 galaxy, providing important information about the emission structure of the surrounding plasma \cite{EventHorizonTelescope:2019dse}. In 2021, the EHT Collaboration released the first polarized images of the M87 black hole, providing valuable information about the magnetic field structure of the surrounding emitting plasma \cite{EventHorizonTelescope:2021bee}.

Since dark matter was first proposed in the 1930s, its nature has remained a mystery. There is some indirect evidence suggesting the existence of dark matter. For example, observed galactic rotation speeds differ from theoretical predictions \cite{Rubin:1970zza}, and the gravitational lensing effect \cite{Vegetti:2023mgp} shows a discrepancy from the theoretical expectations.  Observations of the Bullet Cluster collision \cite{Clowe:2006eq} provide direct empirical evidence: in the Bullet Cluster collision event, since dark matter does not interact electromagnetically, it passes through the collision region without being slowed down, while the baryonic gas is slowed by electromagnetic interactions. As a result, after the collision, the centers of mass of dark matter and baryonic matter become spatially separated.

In the standard cosmological model, dark matter is assumed to consist of non-relativistic, negligible pressure, and collisionless particles, commonly referred to as cold dark matter. N-body simulations of collisionless cold dark matter have revealed characteristic halo density profiles, such as the Navarro–Frenk–White (NFW) profile \cite{Navarro:1995iw}, which exhibits a central cusp in the inner halo region. Another profile derived from observations of dwarf galaxies is the Burkert profile \cite{Burkert:1995yz}, which instead features a core within its scale radius. The discrepancy between theoretical simulations and observational data has led to the well-known ``core-cusp problem \cite{deBlok:2009sp}". Furthermore, CDM simulations predict a large population of low-mass dark matter subhalos around Milky Way-sized galaxies, whereas the number of observed satellite galaxies is considerably smaller. This discrepancy is known as the missing satellites problem \cite{Klypin:1999uc}.
 
In realistic astrophysical environments, black holes are generally not isolated but may reside within dark matter halos. Several models have been proposed regarding black holes surrounded by dark matter. Cardoso developed a model \cite{Cardoso:2021wlq} based on the Einstein cluster approach, assuming dark matter consists of particles uniformly orbiting the black hole in circular motion. Xu's method \cite{Xu:2018wow} treated dark matter as the gravitational source term and obtained the spacetime metric describing a black hole surrounded by dark matter, and has applied it to dark matter spikes \cite{Xu:2021dkv}. Other approaches model dark matter as an anisotropic fluid with only tangential pressure \cite{Jusufi:2022jxu}. Some articles \cite{Gong:2020lev, xie2025equationstatedarkmatter, Zhao:2023itk} have discussed the possible equation of state of dark matter. More recently, various black hole spacetime metrics associated with different dark matter density profiles have been obtained within the Einstein cluster framework \cite{Maeda:2024tsg, Konoplya:2022hbl, Acharyya:2023rnq, Ovgun:2025bol, Liang:2026uut}.
 
The study of black hole accretion disks has a long history, with successive models such as spherically symmetric Bondi accretion \cite{Bondi:1952ni} and thin accretion disk models \cite{Shakura:1973boa, Page:1974he}  having been proposed. In the study of thin disks, the models introduced by Shakura-Sunyaev and Novikov-Thorne \cite{Shakura:1973boa}  played a significant role. Later, alternative and extended models emerged, including slim disks\cite{Abramowicz:1988sp}, thick disks\cite{Abramowicz:1980:Thick}, and ADAFs (Advection-Dominated Accretion Flow) \cite{Narayan:1994xi, Narayan:1994et}, primarily distinguished by different accretion rate regimes, optical depths, and cooling efficiencies.

The accretion disk is a magnetohydrodynamic flow of ionized particles orbiting the black hole. Viscous dissipation within the disk causes it to emit radiation, making it luminous and observable. From the Novikov-Thorne model, energy flux and temperature can be calculated, allowing derivation of spectral luminosity, which can then be compared with observational data.

In this paper, we investigate the impact of dark matter on the observables of thin accretion disks around black holes, using the extended Einstein cluster model proposed by Cardoso \cite{Cardoso:2021wlq} to study black holes embedded in dark matter halos, and comparing them with accretion disks around Schwarzschild black holes. These papers \cite{Heydari-Fard:2022xhr, Heydari-Fard:2026hro, Ranjbar:2025yml, Heydari-Fard:2024wgu, Chowdhury:2025tpt, Patra:2025mnj} have investigated black hole accretion disks within the extended Einstein cluster framework.

In recent years, accretion disks have received considerable attention as astrophysical probes of strong gravitational fields, compact objects, and their surrounding environments. Several studies have investigated the effects of dark matter on accretion disks around black holes \cite{Nieto:2025apz, Boshkayev:2020kle, Pugliese:2022oes, Li:2025ver, Rehman:2025rww}. Within modified theories of gravity, several studies have explored modifications to the energy flux of accretion disks around black holes \cite{Liu:2020vkh, Liu:2021yev, Pun:2008ae, Harko:2009kj, Perez:2017spz}. Accretion disks around other compact objects, such as quark stars and wormholes, have also been investigated \cite{Kovacs:2009gt, Harko:2009xf}. Furthermore, extensive investigations have been carried out on accretion disk dynamics and observational properties in various gravitational backgrounds and astrophysical scenarios \cite {Danila:2015qla, Heydari-Fard:2020ugv, Faraji:2020skq, Chen:2011wb, Lodato:2007hmf, Chen:2011rx, Zhang:2021hit, Fish:2016jil, Muller:2012zw, Gyulchev:2019tvk, Banerjee:2017hzw, Joshi:2013dva, Banerjee:2019sae, Chowdhury:2011aa, Sharipov:2025yfw, Kovacs:2010xm, UmarFarooq:2020aum, Karimov:2018whx, Torres:2002td}.

Beyond accretion disk studies, in particular, numerical simulations of black hole images \cite{Yan:2025mlg, Jusufi:2019nrn} and investigations of the effects of dark matter halos on black hole shadows \cite{Myung:2024tkz,Jusufi:2020cpn} have been conducted. Other studies have explored the properties of periodic orbits, gravitational-wave emission, and axial perturbations of black holes embedded in dark matter halos \cite{Xamidov:2025prl, Haroon:2025rzx, Zhang:2021bdr, Shahzad:2025bft, Shahzad:2025wur, Zhang:2022roh}. In addition, Ref. \cite{Kazempour:2024lcx} investigated accretion processes in the presence of different dark matter density profiles. Meanwhile, studies of dark matter on cosmological scales \cite{Zhang:2023lrz, Hussain:2025uye} have further broadened our understanding of its astrophysical and cosmological implications.

The Dehnen profile\cite{Dehnen1993} was originally introduced as a double power-law, spherically symmetric model for fitting the stellar density distribution of elliptical galaxies, because of its analytic potential and projection formulas. In 2005, this profile was applied to dark matter haloes\cite{Dehnen2005}. This profile represents a family that includes different types such as the Jaffe profile, hollow-core profile, and Hernquist profile. Specifically, we examine the influence of dark matter on the accretion disk when the Jaffe profile is adopted, and this paper gives the metric \cite{Liang:2026uut}. The Jaffe profile originated from a luminous spherically symmetric galaxy model introduced by Jaffe (1983)\cite{Jaffe:1983zz} and was later applied to dark matter halos as an alternative model.

The structure of this paper is as follows: In Section 2, we review the characteristics of dark matter distributions, introduce several commonly used dark matter density profiles, and present the spacetime metric of a black hole surrounded by a dark matter halo within the Einstein cluster framework. In Section 3, we investigate the orbital properties of particles moving along circular equatorial orbits around the black hole. Specifically, we derive the specific energy, specific angular momentum, and effective potential, and determine the innermost stable circular orbit (ISCO). In Section 4, we study the epicyclic frequencies of circular equatorial orbits, focusing on the radial epicyclic frequency and the Keplerian orbital frequency, and examine their dependence on the dark matter halo parameters. In Section 5, within the Novikov–Thorne thin disk framework, we calculate the energy flux, effective temperature, and spectral luminosity of the accretion disk. We then present numerical results to investigate how the dark matter halo mass and scale radius affect the orbital dynamics and radiative properties of the accretion disk. Finally, In Section 6 we summarize our main results and conclusions.

\section{Black hole solutions with dark matter halos}
\renewcommand{\theequation}{2.\arabic{equation}} \setcounter{equation}{0}

In the actual astrophysical environment, black holes are surrounded by dark matter, so the influence of dark matter halos on the spacetime geometry around black holes needs to be considered.
In this work, we choose to adopt the Einstein cluster model developed by Cardoso. In the Einstein cluster model, dark matter particles move along circular timelike geodesics in the spacetime surrounding the black hole. The orbital planes of these particles are randomly oriented, such that the system exhibits spherical symmetry in a statistical sense. Although individual particles are in motion, the overall matter distribution remains static. Since the particles have no radial motion but possess tangential velocities, the resulting energy-momentum tensor has vanishing radial pressure and nonzero tangential pressure. As the radius changes, the density of dark matter also changes.
Therefore $T^{\mu}_{\ \nu} = \operatorname{diag}(-\rho, 0, p_t, p_t)$ \cite{Cardoso:2021wlq}. We consider a static, spherically symmetric metric ansatz and calculate the corresponding Einstein tensor. By substituting the energy-momentum tensor into the Einstein field equations, we can determine the spacetime metric associated with the black hole surrounded by a dark matter halo.  In Cardoso's paper, the static spherically symmetric  metric is  written as
\begin{equation}
ds^2=-f(r)\,dt^2+\frac{dr^2}{g(r)} +r^2\left(d\theta^2+\sin^2\theta\,d\phi^2\right),
\end{equation}
where $g(r)=1-\frac{2m(r)}{r}$. There have been numerous studies on the distribution models of dark matter, such as the NFW, Dehnen, and Burkert profiles. Among them, the NFW profile was obtained from cosmological N-body simulations within the framework of the standard cosmological model. In this framework, dark matter is treated as pressureless, collisionless particles with low velocities, and N-body simulations are performed to obtain the NFW density profile of dark matter. The Burkert profile, in contrast, is an empirical cored density profile introduced to describe the dark matter distributions inferred from the rotation curves of galaxies, particularly dwarf and low-surface-brightness galaxies. The dark matter distribution is constrained by fitting the observed rotation curves and then derives the density distribution formula of dark matter.

The Dehnen profile, introduced by Dehnen \cite{Dehnen1993}, represents a family of analytical density distributions for spherical galaxies and galactic bulges. Unlike models with a fixed central density slope, the Dehnen profile introduces a free parameter $\gamma$ to characterize the inner density distribution. Its density profile is given by Dehnen's dark matter distribution is a collective form of dark matter profiles, multiple dark matter profiles are combined into a parameter expression \cite{Dehnen:1996fa}, 
\begin{equation}
\rho(r) = \frac{(3 - \gamma) M_{\mathrm{halo}}}{4\pi}
\frac{a_{0}}{r^{\gamma} (r + a_{0})^{4 - \gamma}}.
\label{eq: dehnen_density}
\end{equation}
It can be seen that due to the difference in $\gamma$, different dark matter profiles are obtained.
Manjia Liang\cite{Liang:2026uut} extended Cardoso's Hernquist distribution.
When $\gamma$ = 2, which corresponds to the Jaffe profile distribution, the density distribution is

\begin{equation}
\rho(r) = \frac{ M_{\mathrm{halo}}}{4\pi}
\frac{a_{0}}{r^2 (r + a_{0})^2},
\label{eq: dehnen_density}
\end{equation}
 where $M_{\mathrm{halo}}$ denotes the total mass of the dark matter halo and $a_0$ is its characteristic scale radius. The mass function $m(r)$, which describes the mass enclosed within radius $r$, is given by
\begin{equation}
m(r) = M_{\mathrm{BH}} + \frac{M_{\mathrm{halo}} r}{a_{0} + r}
\left(1 - \frac{2M_{\mathrm{BH}}}{r}\right).
\label{eq: jaffe_mass}
\end{equation}
The metric function $f(r)$ is given by \cite{Liang:2026uut}

\begin{equation}
f(r) =  \exp \left[
\frac{2 M_{\mathrm{halo}} \ln \left( \frac{r - 2 M_{\mathrm{BH}}}{a_{0} - 2 M_{\mathrm{BH}} + r} \right)}
{a_{0} + 2 M_{\mathrm{BH}} - 2 M_{\mathrm{halo}}}
\right]
\left(1 - \frac{2 M_{\mathrm{BH}}}{r}\right).
\label{eq: a_gamma2}
\end{equation}

Here, $M_{\rm BH}\equiv M$ denotes the mass of the considered black hole, where $M$ is used throughout
the subsequent analysis.


\section{The Geodesic motion of test particle in black hole in the dark matter halo}
\renewcommand{\theequation}{3.\arabic{equation}} \setcounter{equation}{0}

In this section, we begin to consider the particles on the accretion disk. The particle dynamics are described by the Lagrangian.

\begin{equation}
\mathcal{L} = \frac{1}{2} g_{\mu \nu}\dot{x}^{\mu}\dot{x}^{\nu}. \label{eq:11}
\end{equation}
Here, $g_{\mu\nu}$ denotes the spacetime metric, and an overdot represents differentiation with respect to the proper time $\tau$.
\begin{equation}
ds^{2} = g_{tt}dt^{2} + g_{rr}dr^{2} + g_{\theta \theta}d\theta^{2} + g_{\phi \phi}d\phi^{2}. \label{eq:12}
\end{equation}

Since the metric is independent of the coordinates \(t\) and \(\phi\), the corresponding conjugate momenta are conserved along the geodesics. Using the Euler-Lagrange equations, the corresponding momentum components are obtained as
\begin{align}
p_t &= g_{tt}\dot{t}=-\tilde{E},
\label{eq:13}\\
p_{\phi} &= g_{\phi\phi}\dot{\phi}=\tilde{L},
\label{eq:14}\\
p_r &= g_{rr}\dot{r},
\label{eq:15}\\
p_\theta &= g_{\theta\theta}\dot{\theta}.
\label{eq:16}
\end{align}

Here, $\tilde{E}$ and $\tilde{L}$ denote the conserved specific energy and specific angular momentum of the particle, respectively. Substituting these quantities into the timelike normalization condition,
$g_{\mu\nu}\dot{x}^{\mu}\dot{x}^{\nu}=-1$, we obtain
\begin{equation}
\begin{split}
   g_{rr}\dot{r}^2 + g_{\theta\theta}\dot{\theta}^2 &= -1 - g_{tt}\dot{t}^2 - g_{\phi\phi}\dot{\phi}^2 \\
   &= -1 -\frac{\tilde{E^2}}{g_{tt}}-\frac{\tilde{L^2}}{g_{\phi\phi}}.
\end{split}
\end{equation}

Among them, the effective potential energy is given by the following formula: The effective potential is defined such that the squared specific can be expressed as the sum of the effective potential and the radial kinetic term.
\begin{equation}
V_{\mathrm{eff}}(r) = -g_{tt}\left(1 + \frac{\tilde{L}^2}{g_{\phi\phi}}\right). \label{eq:16}
\end{equation}

On the stable circular orbit in the equatorial plane,   $V_{\mathrm{eff}}(r)=\tilde{E^2} $ and$V_{\mathrm{eff}}'(r)=0 $. By using these conditions, the specific energy, specific angular momentum and angular velocity of the moving particle can be calculated. The details are as follows:

\begin{alignat}{2}
    \tilde{E} &= -\frac{g_{tt}}{\sqrt{-g_{tt} - g_{\phi\phi}\Omega^2}}, &\quad& \label{eq:17}\\
    \tilde{L} &= \frac{g_{\phi\phi}\Omega}{\sqrt{-g_{tt} - g_{\phi\phi}\Omega^2}}, && \label{eq:18}\\
    &\Omega = \sqrt{\frac{-g_{tt, r}}{g_{\phi\phi, r}}}. && \label{eq:19}
\end{alignat}

The innermost stable circular orbit (ISCO) is determined by the marginal stability condition of circular orbits. In particular, the effective potential satisfies
\begin{equation}
V_{\mathrm{eff}}''(r)=0.   \label{eq:20}
\end{equation}

For the static and spherically symmetric spacetime considered in this paper, the metric components are given by \(g_{tt}=-f(r)\) and \((g_{\phi\phi}=r^2)\). Consequently, the ISCO condition can be expressed as
\begin{equation}
2\tilde E^2-\tilde L^2 f''(r)
-\frac{d^2}{dr^2}\left[r^2f(r)\right]=0.     \label{eq:21}
\end{equation}

Tables \ref{tab0} and \ref{ta0} present the orbital quantities and accretion efficiency for different values of the dark matter halo scale radius and mass. Table \ref{tab0} shows that, for a fixed halo mass, increasing the scale radius $a_0$ increases $r_{\mathrm{ISCO}}$, $r_{\mathrm{mb}}$, $E_{\mathrm{ISCO}}^2$, and $\Omega_{\mathrm{ISCO}}^2$, whereas $L_{\mathrm{ISCO}}^2$ and the accretion efficiency decrease. In contrast, Table \ref{ta0} exhibits the opposite trends with increasing halo mass at a fixed scale radius. These results indicate that increasing the halo mass or decreasing its scale radius shifts the ISCO toward smaller radii and enhances the accretion efficiency. For larger-scale radii or smaller halo masses, the orbital quantities gradually approach their corresponding Schwarzschild values.


\begin{table*}[htbp]
    \centering
    \caption{The numerical values of ISCO, marginally bound obit $r_{\mathrm{mb}}$, specific energy, specific angular momentum, specific angular velocity, accretion efficiency for an innermost stable circular orbit for $M_{\mathrm{halo}}/M=10^5$.}\label{tab0}

    \renewcommand{\arraystretch}{1.3}
    \setlength{\tabcolsep}{12pt}

    \begin{tabular*}{\textwidth}{@{\extracolsep{\fill}}cccccccc@{}}
        \hline\hline
        $M_{\mathrm{halo}}/M$ & $a_0/M$ & ISCO& $r_{\mathrm{mb}}$ & $E_{\rm ISCO}^2$ & $L_{\rm ISCO}^2$ & $\Omega_{\rm ISCO}^2$  & $\eta_{\rm NT}$\\
        \hline
        $10^5$ & $10^7$ & 5.89902&3.504448 & 0.671731 & 12.9959 & 0.00382164 & 0.180408 \\
        $10^5$ & $2*10^7$ & 5.94505&3.65258 & 0.768387 & 12.4887 & 0.00419425  & 0.123423 \\
        $10^5$ & $10^8$ & 5.98821&3.89009 & 0.860781 & 12.0963 & 0.00452771  & 0.0722171 \\
        $10^5$ & $10^{10}$ & 5.99988&3.99838 &0.888522 &12.001 & 0.00462818 &0.0573855 \\
        --- &---& 6&4& 0.888889& 12& 0.00462963&0.057191\\
        \hline\hline
    \end{tabular*}
\end{table*}
\vspace{0.5cm}

\begin{table*}[htbp]
    \centering
    \caption{The numerical values of ISCO, marginally bound obit $r_{\mathrm{mb}}$, specific energy, specific angular momentum, specific angular velocity,  accretion efficiency for an innermost stable circular orbit for \(a_{0}/M=10^{10}\).}\label{ta0}

    \renewcommand{\arraystretch}{1.3}
    \setlength{\tabcolsep}{12pt}

    \begin{tabular*}{\textwidth}{@{\extracolsep{\fill}}cccccccc@{}}
        \hline\hline
        \(M_{\mathrm{halo}}/M\) & \(a_0/M\) & ISCO & $r_{\mathrm{mb}}$ & \(E_{\rm ISCO}^2\) & \(L_{\rm ISCO}^2\) & \(\Omega_{\rm ISCO}^2\) & $\eta_{\rm NT}$ \\
        \hline
        $10^5$ & $10^{10}$ & 5.99988 &3.99838& 0.888522 & 12.001 & 0.004628  & 0.0573855 \\
        $10^7$ & $10^{10}$ & 5.98821 &3.86263& 0.852874 & 12.0963 & 0.004486  & 0.0764884 \\
        $5*10^7$ & $10^{10}$ & 5.94505 &3.56483& 0.721636 & 12.4887 & 0.003939  & 0.150509 \\
        $10^8$ & $10^{10}$ & 5.89902 &3.39564&0.583406 &12.9959 & 0.003319 &0.23619 \\
        --- &---& 6&4& 0.888889& 12& 0.00462963&0.057191\\
        \hline\hline
    \end{tabular*}
\end{table*}
\vspace{0.5cm}
\section{Epicyclic Frequencies of Circular Equatorial Orbits}
In accretion disks, restoring forces give rise to radial and vertical oscillations when particles are perturbed from their equilibrium positions. A small radial displacement of a particle from a stable circular orbit causes it to oscillate about its original radius under the influence of the restoring force. These oscillations are characterized by the radial epicyclic frequency $\Omega_r$. Likewise, a small perturbation perpendicular to the equatorial plane results in vertical oscillations with frequency $\Omega_\theta$. Together with the orbital angular frequency $\Omega_K$, these frequencies describe the motion of particles near circular equatorial orbits and provide useful information about the dynamical properties of the accretion disk. In the present study, we investigate how the surrounding dark matter halo influences these characteristic frequencies through its modification of the black hole spacetime.
To determine the radial and vertical epicyclic frequencies, we consider small perturbations of circular geodesic motion in the equatorial plane. The corresponding radial and vertical effective potentials are introduced as $\frac{1}{2}(dr/dt)^2=V_{\mathrm{eff}}^{(r)}$ and $\frac{1}{2} (d\theta/dt)^2=V_{\mathrm{eff}}^{(\theta)}$, respectively. For purely radial perturbations, we impose $u^\theta=0$, while for vertical perturbations, we set $u^r=0$. By considering $u^r=dr/d\tau=(dr/dt)u^t$ and $u^\theta=d\theta/d\tau=(d\theta/dt)u^t$, along with the conserved specific energy $E$ and angular momentum $L$, we have 
 \begin{equation}
\frac{1}{2}\left(\frac{dr}{dt}\right)^2 = -\frac{f^2(r)g(r)}{2\tilde{E}^{\,2}} \left[
1-\frac{\tilde{E}^{\,2}}{f(r)} +\frac{\tilde{L}^{\,2}}{r^2} \right] =V_{\mathrm{eff}}^{(r)}.
\label{d2}
\end{equation}
\begin{equation}
\frac{1}{2}\left(\frac{d\theta}{dt}\right)^2 = -\frac{f^2(r)}{2r^2\tilde{E}^{\,2}} \left[
1-\frac{\tilde{E}^{\,2}}{f(r)} +\frac{\tilde{L}^{\,2}}{r^2\sin^2\theta} \right] =V_{\mathrm{eff}}^{(\theta)}.
\label{d3}
\end{equation}

We assume that the particle undergoes small perturbations from its circular equatorial orbit, specified by $r=r_{0}$ and $\theta=\theta_{0}=\pi/2$. The corresponding deviations along the radial and vertical directions are denoted by $\delta r=r-r_{0}$ and $\delta\theta=\theta-\theta_{0}$, respectively. For small perturbations, the equations of motion can be linearized in the vicinity of the equilibrium orbit. The radial equation is then obtained by taking the derivative of the corresponding effective-potential equation with respect to the coordinate time $t$ expressed as
\begin{eqnarray}
\frac{d^2 r}{dt^2}=\frac{d V_{eff}^{(r)}}{dr}.\label{d3}
\end{eqnarray}
At the circular orbit, the first derivative of the effective potential vanishes, and the second derivative of the radial effective potential governs the radial motion under small perturbations. Therefore, we obtain
\begin{eqnarray}
\frac{d^2}{dt^2} (\delta r)=\frac{d^2 V_{eff}^{(r)}}{dr^2} (\delta r) \Rightarrow (\delta \ddot{r})+\Omega_{r}^2(\delta r)=0.\label{d4}
\end{eqnarray}
Here $\Omega_{r}^{2}\equiv -d^{2}V_{\mathrm{eff}}^{(r)}/dr^{2}$ indicates the squared radial epicyclic frequency, and $\delta r$ represents a small radial displacement from the reference circular orbit, while a dot indicates differentiation with respect to the coordinate time $t$. Similarly, for the vertical perturbation $\delta\theta=\theta-\theta_{0}$, where $\theta_{0}=\pi/2$, the linearized equation of motion takes the form
\begin{eqnarray}
\frac{d^2 (\delta\theta)}{dt^2}=\frac{{d^2 V_{eff}}^{(\theta)}}{d\theta^2} (\delta\theta) \Rightarrow (\delta \ddot{\theta})+\Omega_{\theta}^2(\delta \theta)=0.\label{d5}
\end{eqnarray}
In the above expression, $\Omega_{\theta}^{2}\equiv-d^{2}V_{\mathrm{eff}}^{(\theta)}/d\theta^{2}$ represents the squared vertical epicyclic frequency. By solving the radial perturbation equation for circular orbits in the equatorial plane, the squared radial epicyclic frequency is obtained as
\begin{equation}
\Omega_r^2= \frac{g(r)}{2} \left[ f''(r) -\frac{2[f'(r)]^2}{f(r)} +\frac{3f'(r)}{r} \right].
\label{eq: radial_epicyclic_standard}
\end{equation}
Also, the associated squared vertical epicyclic frequency is
\begin{equation}
\Omega_{\theta}^2 = \frac{f^2(r)L^2}{E^2r^4}.
\label{d7}
\end{equation}

The Keplerian orbital frequency for the considered
static and spherically symmetric spacetime is given by
\begin{equation}
\Omega_K \equiv \frac{d\phi}{dt} = \frac{\dot{\phi}}{\dot{t}} =
\sqrt{\frac{f'(r)}{2r}}.
\label{eq:Omega_K}
\end{equation}
Consequently, the vertical epicyclic frequency coincides with the Keplerian orbital frequency,
\begin{equation}
\Omega_{\theta}^2 = \Omega_K^2 = \frac{f'(r)}{2r}.
\label{eq:Omega_theta}
\end{equation}

\begin{figure*}
\centering
\includegraphics[scale=0.65]{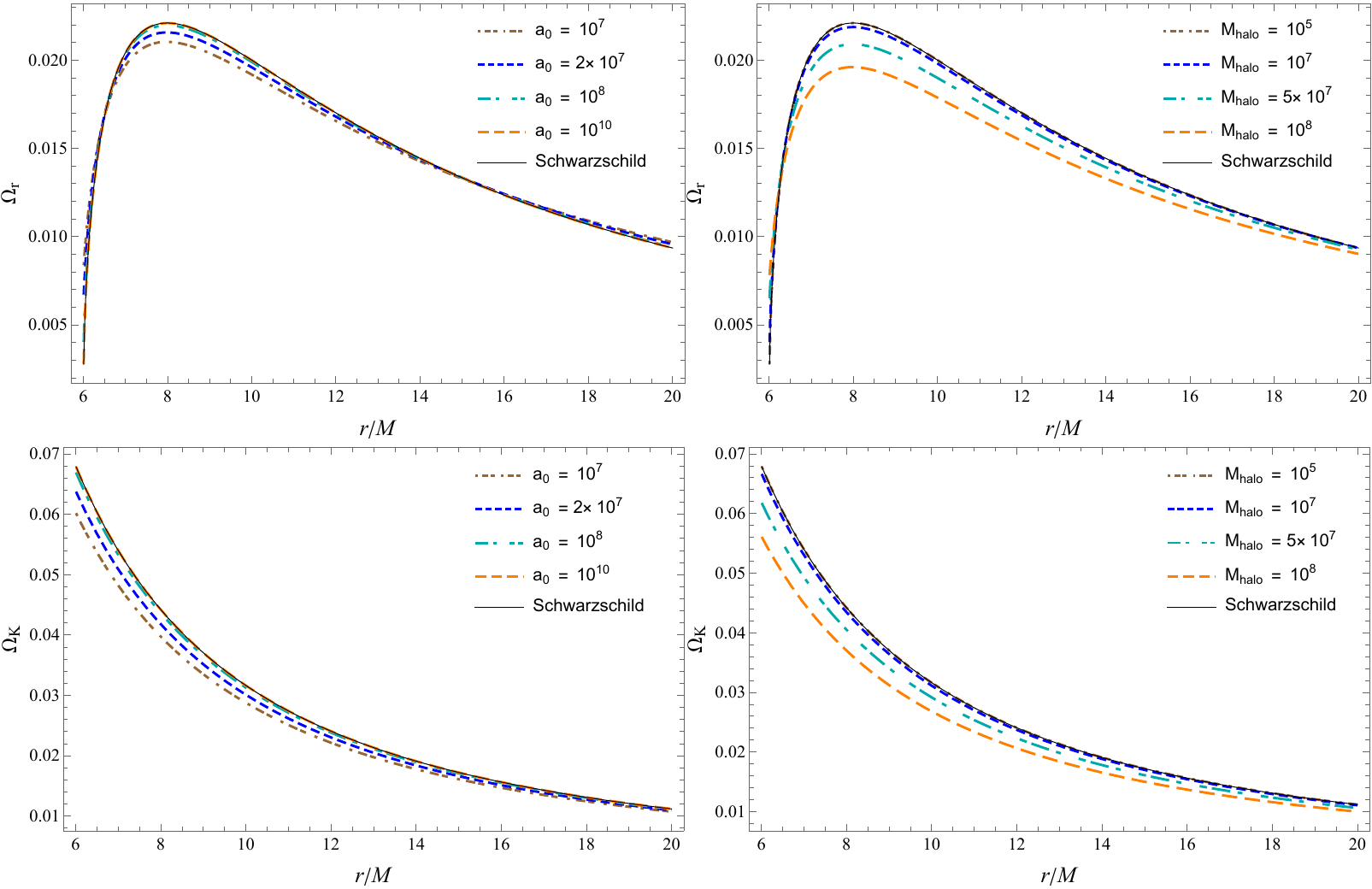}
\caption{{Left: Radial epicyclic frequency $\Omega_r$ (upper panels) and Keplerian orbital frequency $\Omega_K$ (lower panels) as functions of the radial coordinate $r/M$ for a Schwarzschild BH surrounded by a dark matter halo. The left panels show the effects of varying the halo scale radius $a_0=10^7$, $2\times10^7$, $10^8$, and $10^{10}$ at fixed $M_{\rm halo}=10^5$, whereas the right panels illustrate the influence of the halo mass $M_{\rm halo}=10^5$, $10^7$, $5\times10^7$, and $10^8$ at fixed $a_0=10^{10}$. The Schwarzschild solution is included for comparison, with $M=1$.}\label{fig: epi}}
\label{Temperature}
\end{figure*}

Figure~\ref{fig: epi} shows the radial epicyclic frequency $\Omega_r$ and Keplerian orbital frequency $\Omega_K$ for different values of the dark matter halo scale radius $a_0$ and mass $M_{\rm halo}$. The upper panels represent the behavior of $\Omega_r$, which initially rises with radial distance, reaches a maximum near $r/M\simeq8$, and subsequently declines at larger radii. The upper-left panel shows that increasing $a_0$ enhances the radial epicyclic frequency near its maximum, bringing it closer to the Schwarzschild result. The deviation is more pronounced for smaller scale radii, particularly for $a_0=10^7$, whereas the frequency profile for $a_0=10^{10}$ is nearly indistinguishable from the Schwarzschild case. In contrast, the upper-right panel shows that rising the value of $M_{\rm halo}$ reduces the radial epicyclic frequency, with the largest deviation occurring for $M_{\rm halo}=10^8$. The differences between the frequency profiles are more noticeable near the maximum and become smaller at larger radial distances. 

The lower panels of Fig.~\ref{fig: epi} show the variation of the Keplerian orbital frequency $\Omega_K$ with the radial coordinate $r/M$ for different values of the dark matter halo scale radius $a_0$ and mass $M_{\rm halo}$. In both cases, $\Omega_K$ decreases monotonically with increasing radial distance. The lower-left panel shows that increasing $a_0$ raises the orbital frequency, approaching the Schwarzschild result for larger scale radii. In particular, the frequency profile for $a_0=10^{10}$ is nearly indistinguishable from the Schwarzschild case, while the largest deviation occurs for $a_0=10^7$. In contrast, the lower-right panel indicates that increasing $M_{\rm halo}$ reduces $\Omega_K$, with the most significant reduction observed for $M_{\rm halo}=10^8$. The variations in frequency profiles are more prominent in the inner region and gradually diminish as radial distance rises. This shows that the influence of the dark matter halo on the orbital frequency becomes more prominent for smaller halo scale radii and larger halo masses.


\section{Observational properties of thin accretion disks}
This section adopt the steady-state thin accretion disk model developed by Novikov and Thorne (NT) \cite{Shakura:1973boa,Page:1974he} to analyze the impact of dark matter halos on the accretion properties of the black hole under consideration. The disk is assumed to be geometrically thin, with its vertical thickness ($h$) much smaller than its radial extent ($h\ll r$). It lies in the equatorial plane, where the accreting matter moves along nearly circular Keplerian orbits outside the innermost stable circular orbit (ISCO), which defines the inner edge of the disk. The background spacetime is assumed to be static and spherically symmetric, while the accretion flow is stationary and axisymmetric, with a constant mass accretion rate. In the NT accretion disk model, the energy-momentum tensor of the accreted matter can be written as

\begin{equation}
T^{\mu \nu} = \rho_0 u^\mu u^\nu + 2u^{(\mu}q^{\nu)} + t^{\mu \nu}. \label{eq:21}
\end{equation}
The quantities \(\rho_0\), \(u^\mu\), \(q^\mu\), and \(t^{\mu\nu}\) denote the rest-mass density, four-velocity, heat-flow vector, and viscous stress tensor, respectively.

By applying the conservation laws of rest mass, energy, and angular momentum to the accretion flow, the fundamental equations governing the disk can be derived within the NT framework.
\begin{align}
\dot{M}_0 = -2\pi \sqrt{-g}\Sigma u^r = \mathrm{const}, \label{eq:22}\\
\dot{M}_0\tilde{E} -2\pi \sqrt{-g}\Omega W_\phi^r ]_{, r} = 4\pi rF(r)\tilde{E}, \label{eq:23}\\
\dot{M}_0\tilde{L} -2\pi \sqrt{-g} W_\phi^r ]_{, r} = 4\pi rF(r)\tilde{L}. \label{eq:24}
\end{align}
Among them, $g$ is the metric determinant, and the time-averaged surface density is given by the following equation,
\begin{align}
    \Sigma(r) = \int_{-h}^{h}\langle \rho_0\rangle dz, \label{eq:25}\\
    W_{\phi}^{r} = \int_{-h}^{h}\langle t_{\phi}^{r}\rangle dz. \label{eq:26}
\end{align}

Here, $z$ represents the cylindrical coordinate. Through the calculation of the equation, the average energy flux\cite{Shakura:1973boa} emitted by the accretion disk is obtained:
\begin{equation}
F(r) = -\frac{\dot{M}_0\Omega_{, r}}{4\pi\sqrt{-g}\left(\tilde{E} - \Omega\tilde{L}\right)^2}\int_{r_{\mathrm{ISCO}}}^{r}\left(\tilde{E} -\Omega\tilde{L}\right)\tilde{L}_{, r}\, dr. \label{eq:27}
\end{equation}
\begin{figure*}[!t]
\centering
\includegraphics[scale=0.95]{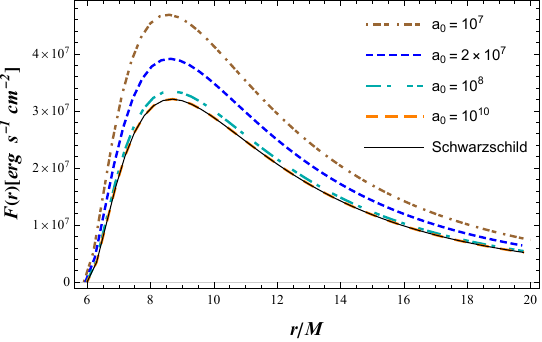}
\includegraphics[scale=0.95]{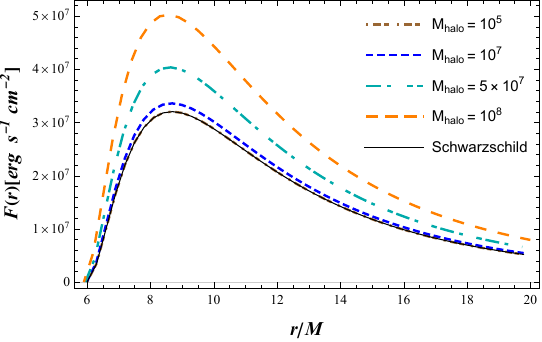}
\caption{\textbf{Left:} Energy flux distribution of the accretion disk around a black hole of mass \( M = 10^{6} \, M_{\odot} \) with \( M_{\mathrm{halo}} = 10^5 M \), for different dark matter profile scale radii \( a_0 \). \textbf{Right:} Energy flux curves at a fixed scale radius \( a_{0} = 10^{10} M \), for various dark matter halo masses.}
\label{flux}
\end{figure*}
 \begin{figure*}[!t]
\centering
\includegraphics[scale=0.95]{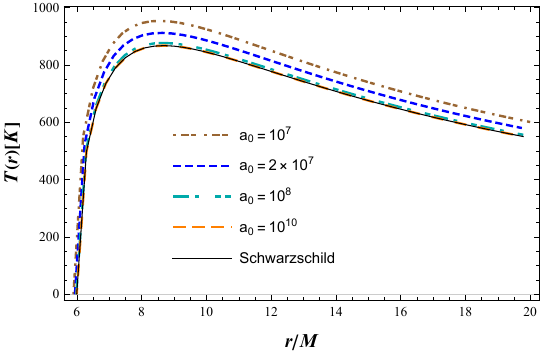}
\includegraphics[scale=0.95]{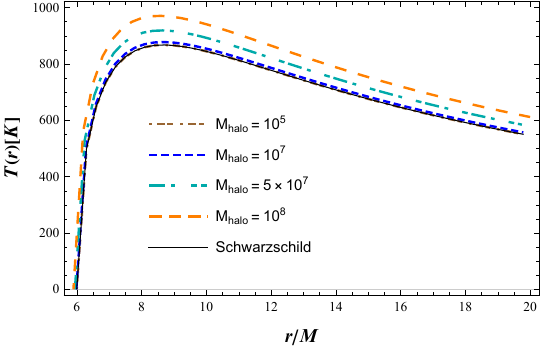}
\caption{\textbf{Left:} Temperature distribution of the accretion disk around a black hole of mass \( M = 10^{6} \, M_{\odot} \) with \( M_{\mathrm{halo}} = 10^5 M \), for different dark matter profile scale radii \( a_0 \). \textbf{Right:} temperature curves at a fixed scale radius \( a_{0} = 10^{10} M \), for various dark matter halo masses.}
\label{Temp}
\end{figure*}

Figure \ref{flux} represents the influence of the dark matter halo parameters on the energy flux emitted from the accretion disk. The left panel indicates that, for a fixed halo mass, decreasing the scale radius \(a_0\) increases the energy flux, particularly in the disk's inner region. This behavior arises because a smaller scale radius corresponds to a more concentrated dark matter distribution, which modifies the spacetime geometry and orbital properties of the accreting matter. Similarly, the right panel shows that rising the halo mass produces a higher energy flux. For larger-scale radii or smaller halo masses, the corresponding curves approach the Schwarzschild solution, indicating a weaker influence of the dark matter halo.

Since the steady-state thin disk model assumes thermodynamic equilibrium, the disk surface of the accretion disk can be regarded as an ideal blackbody radiator. Therefore, the relationship between the temperature and energy flux of the accretion disk can be expressed by the following equation. It can also be calculated from the Planck blackbody radiation formula,
\begin{equation}
F(r) = \sigma_{\mathrm{SB}}T^4(r). \label{eq:28}
\end{equation}

Figure  \ref{Temp} shows the radial temperature distribution of the accretion disk for different dark matter halo parameters. From the figure, we notice that the temperature initially rises with radial distance, reaches a maximum near the inner region of the disk, and then gradually declines at larger radii. The left panel illustrates that as the value of the scale radius $a_{0}$ declines, the temperature of the disk rises. This is because a smaller scale radius corresponds to a higher concentration of dark matter near the BH, which enhances the energy flux and consequently the disk temperature. Similarly, the right panel shows that rising the halo mass leads to a higher temperature. For larger-scale radii or smaller halo masses, the profiles of the temperature approach the Schwarzschild case, representing that the impact of dark matter becomes less significant.

\begin{figure*}[!t]
\centering
\includegraphics[scale=0.4]{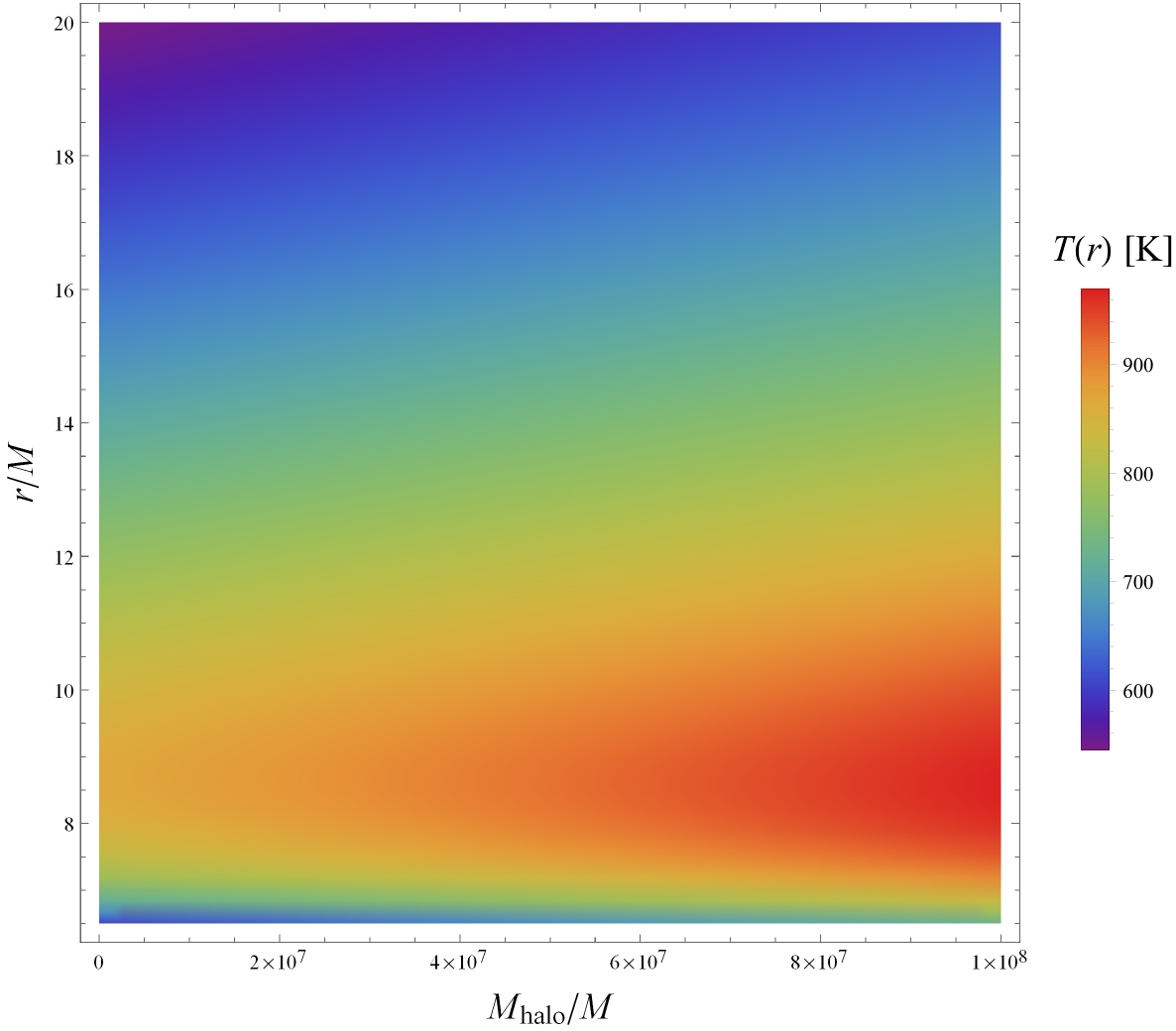}
\includegraphics[scale=0.4]{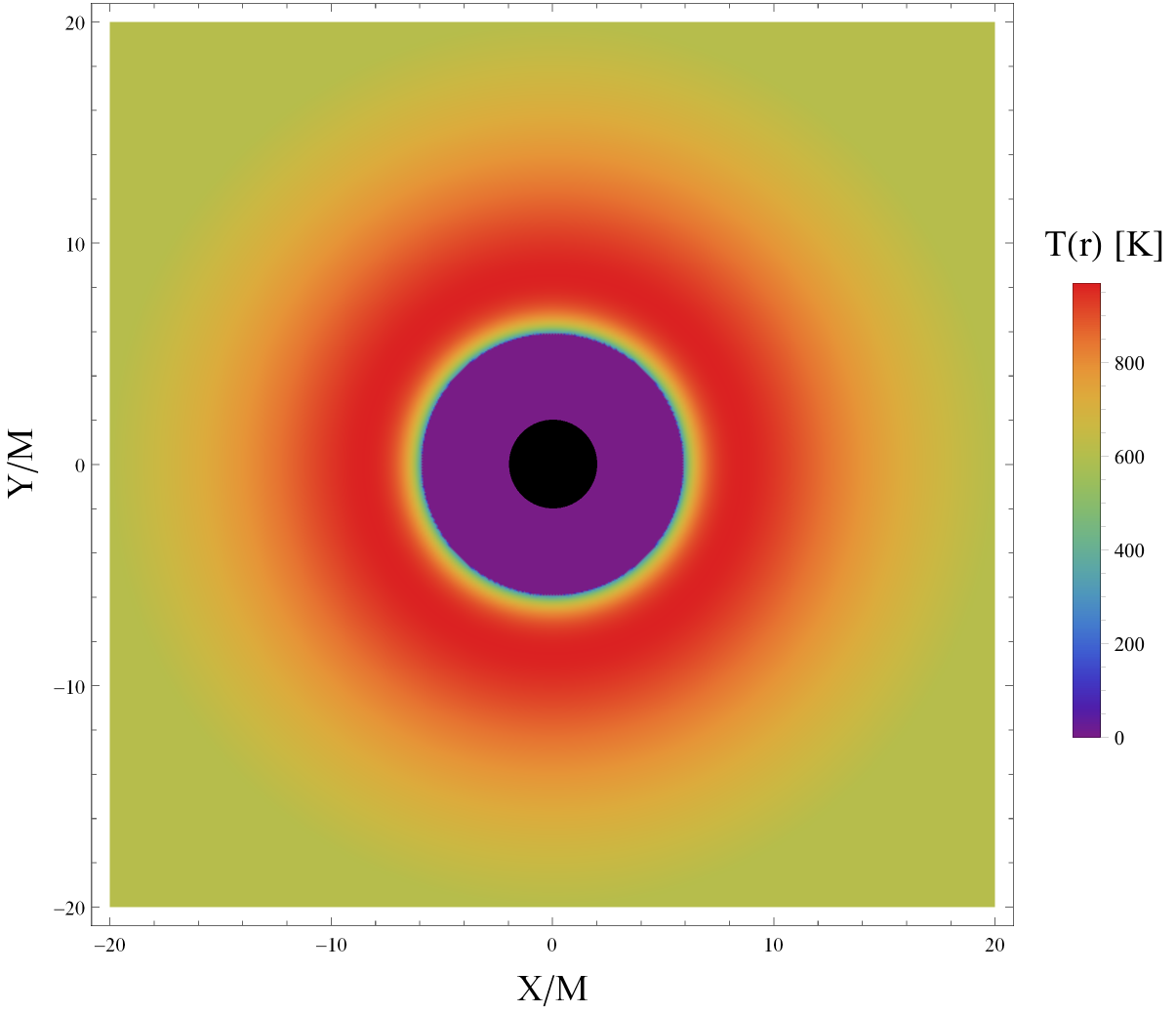}
\caption{Temperature distribution of the accretion disk in the presence of a dark-matter halo. Left: $T(r)$ as a function of $M_{\rm halo}/M$ and $r/M$. Right: temperature map in the equatorial $(X, Y)$. The black central region denotes the black hole horizon, and the ISCO sets the inner edge of the disk.}
\label{Temperature}
\end{figure*}

\begin{figure*}[!t]
\centering
\includegraphics[scale=0.4]{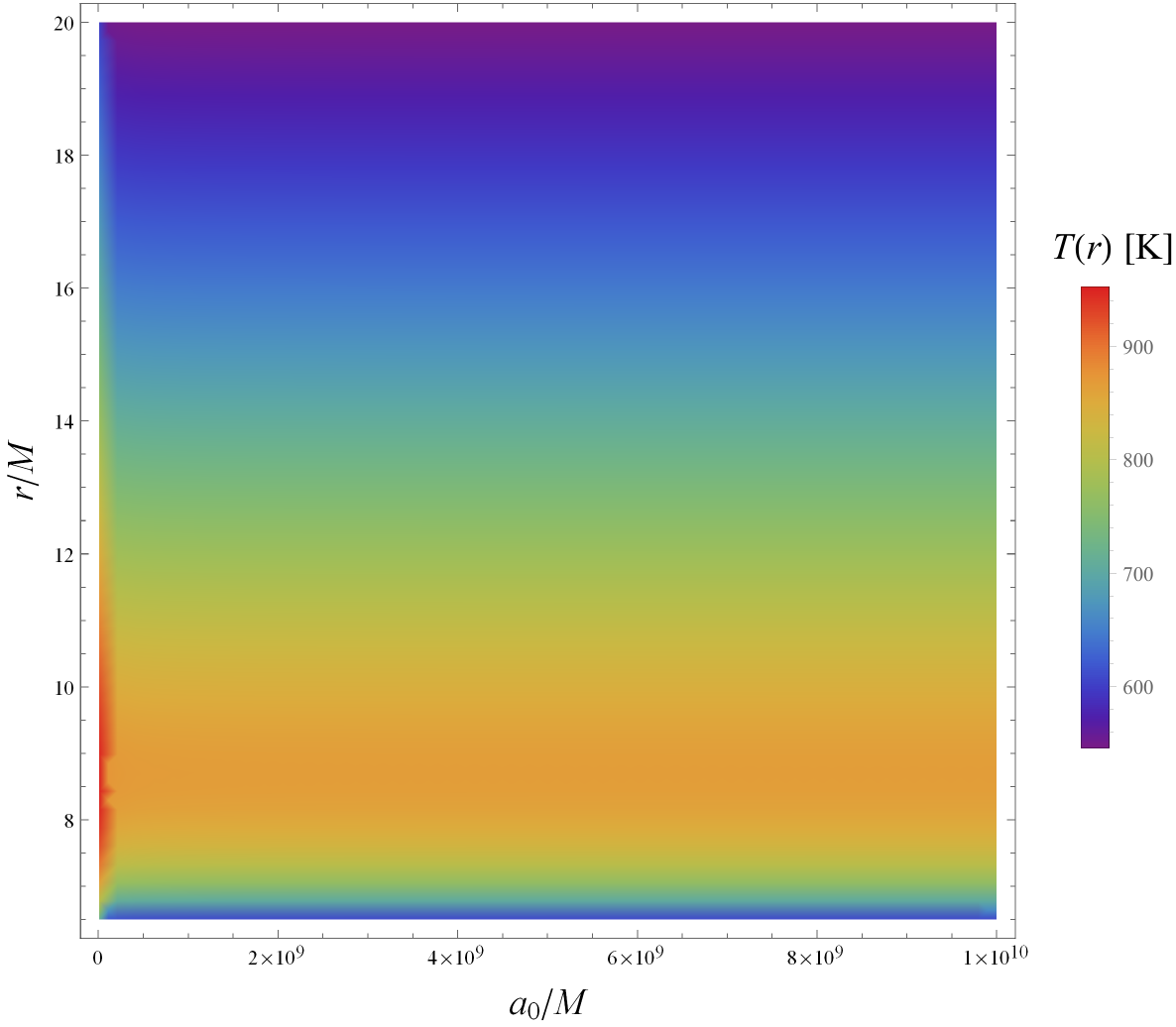}
\includegraphics[scale=0.4]{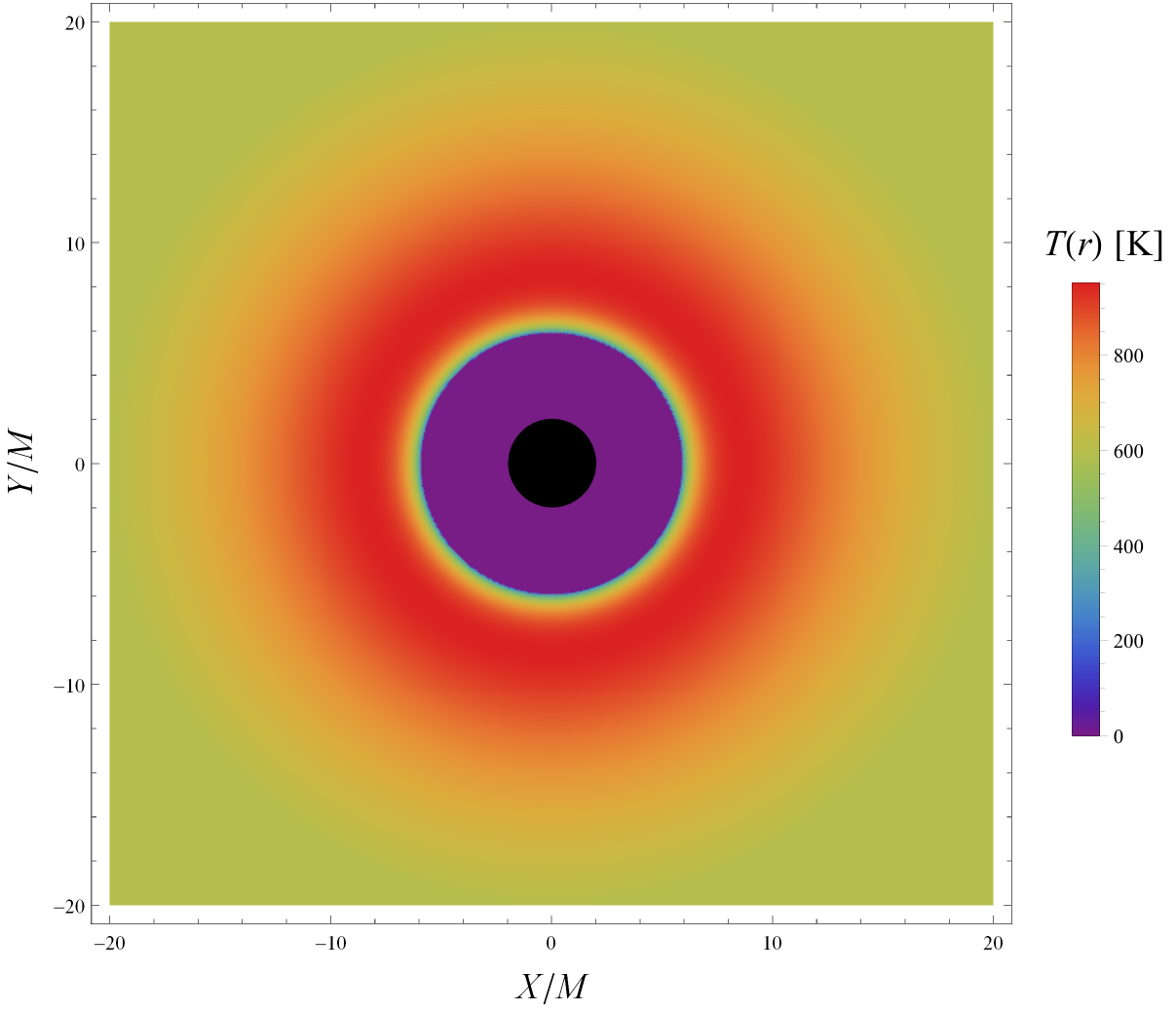}
\caption{Temperature distribution of the accretion disk in the presence of a dark matter halo. Left: $T(r)$ as a function of $a_0/M$ and $r/M$. Right: Temperature distribution in the equatorial $(X,Y)$ plane. The central black region represents the black hole horizon, while the ISCO determines the inner edge of the accretion disk.}
\label{Temperature1}
\end{figure*}
Figures \ref{Temperature} and \ref{Temperature1} indicate the temperature distribution of the thin accretion disk for different dark matter halo parameters. The left panels show the temperature contours in the \((M_{\rm halo}/M, r/M)\) and \((a_0/M, r/M)\) planes, respectively, while the right panels present the corresponding two-dimensional temperature maps. Figure \ref{Temperature} shows that rising the halo mass enhances the disk temperature. In contrast, Figures \ref{Temperature1} indicates that the temperature declines as the halo scale radius increases. The two-dimensional maps show that the highest temperatures are observed in an annular region near the inner edge of the disk, followed by a gradual decrease toward larger radii. These results indicate that a more massive and compact dark matter halo produces a stronger influence on the thermal properties of the accretion disk.
Here, $\sigma_{\mathrm{SB}}$ represents the Stefan-Boltzmann constant. As follows: The observed spectral luminosity is expressed by the following equation\cite{Torres:2002td}:

\begin{figure*}[!t]
\centering
\includegraphics[scale=0.95]{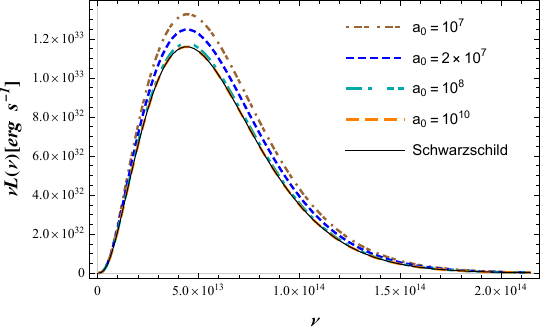}
\includegraphics[scale=0.95]{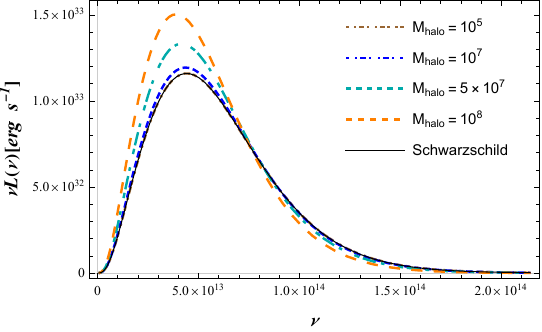}
\caption{\textbf{Left:}spectral luminosity distribution of the accretion disk around a black hole of mass \( M = 10^{6} \, M_{\odot} \) with \( M_{\mathrm{halo}} = 10^5 M \), for different dark matter profile scale radii \( a_0 \). \textbf{Right:}spectral luminosity curves at a fixed scale radius \( a_{0} = 10^{10} M \), for various dark matter halo masses.}
\label{lum}
\end{figure*}

\begin{table*}[htbp]
    \centering
    \caption{The numerical results for the ISCO,
the maximum of energy flux, maximum of temperature,
the critical frequency, maximum emission spectrum,
and efficiency of an accretion disk  with $M_{\mathrm{halo}}=10^{5}M$ and different values of $a_0$. The last row corresponds to the Schwarzschild BH without dark matter.}\label{tab}

    \renewcommand{\arraystretch}{1.3}
    \setlength{\tabcolsep}{12pt}

    \begin{tabular*}{\textwidth}{@{\extracolsep{\fill}}cccccc@{}}
        \hline\hline
        $M_{\rm halo}/M$
& $a_0/M$
& \makecell{$F_{\max}(r)$\\
$[\mathrm{erg\, s^{-1}cm^{-2}}]\times10^{7}$}
& \makecell{$T_{\max}(r)$\\
$[\mathrm{K}]\times10^2$}
& \makecell{$\nu_c$\\
$[\mathrm{Hz}]\times10^{13}$}
& \makecell{$\nu L(\nu)_{\max}$\\
$[\mathrm{erg}]\times10^{33}$} \\
        \hline
       $10^5$ & $10^7$ & 4.68399 & 9.5239 & 4.55577 & 1.32528 \\

$10^5$ & $2*10^7$ & 3.91184 & 9.11337 & 4.55577 & 1.24658 \\

$10^5$ & $10^8$ & 3.34365 & 8.76288 & 4.33358 & 1.17598 \\

$10^5$ & $10^{10}$ & 3.20547 & 8.67094 & 4.33358 & 1.15876 \\

--- & --- & 3.20395 & 8.6699 & 4.33358 & 1.15863 \\

        \hline\hline
    \end{tabular*}
\end{table*}
\vspace{0.5cm}

\begin{table*}[htbp]
    \centering
    \caption{The numerical results for $r_{\mathrm{ISCO}}$, the maximum of energy flux, maximum of temperature, the critical frequency, maximum emission spectrum, and efficiency of an accretion disk with $a_0=10^{10}M$ and different values of $M_{halo}$. The last row corresponds to the Schwarzschild BH without dark matter.}\label{tab1}

    \renewcommand{\arraystretch}{1.3}
    \setlength{\tabcolsep}{12pt}

    \begin{tabular*}{\textwidth}{@{\extracolsep{\fill}}cccccc@{}}
        \hline\hline

$M_{\rm halo}/M$
& $a_0/M$
& \makecell{$F_{\max}(r)$\\
$[\mathrm{erg\, s^{-1}cm^{-2}}]\times10^{7}$}
& \makecell{$T_{\max}(r)$\\
$[\mathrm{K}]\times10^2$}
& \makecell{$\nu_c$\\
$[\mathrm{Hz}]\times10^{13}$}
& \makecell{$\nu L(\nu)_{\max}$\\
$[\mathrm{erg}]\times10^{33}$}\\
        \hline
   $10^5$ & $10^{10}$ & 3.20535 & 8.67094 & 4.33358 & 1.15898 \\

$10^7$ & $10^{10}$ & 3.35894 & 8.77299 & 4.33358& 1.19357 \\

$5*10^7$ & $10^{10}$ & 4.03608 &9.18517 & 4.12223 & 1.33113 \\

$10^8$ & $10^{10}$ & 5.02485 & 9.70238 & 3.92118 & 1.5026 \\

--- & --- & 3.20382 & 8.6699 & 4.33358&1.15863 \\

        \hline\hline
    \end{tabular*}
\end{table*}
\vspace{0.5cm}
\begin{equation}
L(\nu) = \frac{8\pi h\cos\gamma}{c^2}\int_{r_{\mathrm{in}}}^{r_{\mathrm{out}}}\int_{0}^{2\pi}\frac{\nu_{\mathrm{e}}^3 r\, dr\, d\phi}{\exp\left[\dfrac{h\nu_{\mathrm{e}}}{k_{\mathrm{B}}T}\right] - 1}. \label{eq:29}
\end{equation}
This equation represents the spectral luminosity of the accretion disk as observed by the observer. Here, $g$ is the redshift factor. Assuming that the disk emits locally as a blackbody, the observed spectral luminosity can be obtained by integrating the redshifted blackbody radiation over the entire disk surface. $r_{\rm in}$ is the inner radius of the accretion disk, $r_{\rm out}$ is the outer radius, and $\gamma$ is the inclination angle of the accretion disk. The observed spectral flux is multiplied by $4\pi d^2$, where $d$ is the distance between the observer and the black hole, to obtain the isotropic-equivalent spectral luminosity of the accretion disk. The redshift factor is given by the following formula.

\begin{equation}
1 + z = \frac{1 + \Omega r \sin \phi \sin \gamma}{\sqrt{-g_{tt} - \Omega^2 g_{\phi\phi}}}. \label{eq:30}
\end{equation}

Accretion efficiency is an important physical quantity characterizing the radiative efficiency of an accretion disk. In the Novikov–Thorne (NT) model, assuming a vanishing torque at the ISCO, the accretion efficiency is defined as

\begin{equation}
\eta_{\rm NT}=1-E_{\rm ISCO}.
\end{equation}

Here, $E_{\rm ISCO}$ is the specific energy of a particle at the innermost stable circular orbit (ISCO) around the black hole. Under the standard assumptions of the NT model, the gravitational binding energy released by the accreting matter outside the ISCO is radiated away, whereas the matter inside the ISCO rapidly plunges into the black hole, with its radiative contribution usually neglected.

Figure \ref{lum} shows the spectral luminosity of the accretion disk for different dark matter halo parameters. The left panel shows that decreasing the halo scale radius enhances the peak spectral luminosity, while the right panel indicates that increasing the halo mass produces a similar effect. In both cases, the isotropic-equivalent spectral luminosity initially increases with frequency, reaches a maximum, and subsequently decreases at higher frequencies. For larger-scale radii or smaller halo masses, the curves approach the Schwarzschild solution, indicating a weaker influence of the dark matter halo on the emitted radiation.

Tables \ref{tab} and \ref{tab1} summarize the numerical results for the ISCO radius, maximum energy flux, maximum temperature, critical frequency, and peak spectral luminosity for different dark matter halo parameters. Table \ref{tab} shows that, for a fixed halo mass, decreasing the scale radius shifts the ISCO inward and increases the maximum energy flux, temperature, and spectral luminosity. Similarly, Table \ref{tab1} indicates that increasing the halo mass enhances these radiative quantities. For larger scale radii or smaller halo masses, the results approach the Schwarzschild values, indicating that the influence of the dark matter halo becomes less significant.

\section{Conclusion}

In this work, we investigated the properties of a thin accretion disk around a Schwarzschild black hole immersed in a dark matter halo, using the Einstein cluster model and the Novikov–Thorne accretion disk framework. We considered a black hole of mass \(M=10^{6}M_{\odot}\) and a mass accretion rate of \(\dot{M}_{0}=10^{-12}M_{\odot}/\mathrm{yr}\). We examined the influence of the dark matter halo mass and scale radius on circular particle motion, epicyclic frequencies, the innermost stable circular orbit (ISCO), and the radiative properties of the accretion disk.

To assess the astrophysical relevance of the adopted parameters, we considered the Milky Way as an example. The black hole at the center of the Milky Way is approximately $4.3*10^6$ times the mass of the Sun \cite{GRAVITY:2021xju}. Based on the axisymmetric Jeans model, the mass of the dark matter halo in the Milky Way is approximately \(M_{200} = 0.820^{+0.210}_{-0.186} \times 10^{12} M_{\odot}\)\cite{Zhang:2025abc}.  The measured scale radius of the NFW halo of the Milky Way galaxy shows significant dispersion, ranging from approximately 3.6 kpc based on Gaia DR3\cite{Lim:2023lss} to approximately 12.5 kpc based on large-scale rotation curves\cite{Sofue:2011kw}. The mass of dark matter in real galaxies is quite large. The dark matter mass of the Milky Way is approximately $10^5$ times that of a black hole, and its scale radius is about $10^{10}-10^{11}$ times. Therefore, we use  $a_0=10^{10}M$  $M_{\mathrm{halo}}=10^{5}M$ to further investigate the dependence of the disk properties on the dark matter parameters, and we also considered a broader range of parameter values, including configurations with smaller scale radii and larger halo masses. 

We first investigated the orbital properties of particles moving along circular equatorial orbits. The results show that decreasing the halo scale radius or increasing its mass shifts the ISCO toward smaller radii. The marginally bound orbit and the specific energy at the ISCO also decrease, whereas the specific angular momentum and accretion efficiency increase. For larger halo scale radii or smaller halo masses, these quantities approach their corresponding Schwarzschild values.

We also examined the radial epicyclic frequency \(\Omega_r\) and the Keplerian orbital frequency \(\Omega_K\) for different values of the dark matter halo parameters. The radial epicyclic frequency initially increases with radial distance, reaches a maximum, and subsequently decreases, while the Keplerian orbital frequency decreases monotonically with increasing radius. Increasing the halo mass reduces both frequencies, whereas increasing the halo scale radius brings their profiles closer to the Schwarzschild results. The differences are more noticeable in the inner region and gradually become smaller at larger radial distances.

The energy flux, disk temperature, and emission spectrum were calculated within the Novikov–Thorne model. For a fixed halo mass, decreasing the scale radius increases the energy flux and disk temperature, particularly near the inner region of the accretion disk. Similarly, increasing the halo mass at a fixed scale radius enhances these quantities. The temperature distributions in the equatorial plane show that the highest temperatures occur in an annular region near the inner edge of the disk, followed by a gradual decrease toward larger radii.

The spectral luminosity exhibits a similar dependence on the halo parameters. Decreasing the scale radius or increasing the halo mass enhances the peak spectral luminosity. The numerical results also indicate that the accretion efficiency increases as the dark matter halo becomes more massive or more concentrated. These changes are accompanied by an inward displacement of the ISCO, indicating that the surrounding dark matter halo modifies both the orbital motion and the radiative properties of the accretion disk.

For halo parameters corresponding to large characteristic scale radii and relatively weak modifications of the black hole spacetime, the calculated quantities remain close to their Schwarzschild values. The influence of the dark matter halo becomes more pronounced when its scale radius decreases or its mass increases. Thus, within the considered model, the dark matter halo can modify the characteristic orbital frequencies, the location of the inner disk boundary, the accretion efficiency, and the emitted radiation, with the strongest effects arising for more compact or massive halos.

 \section*{Acknowledgements}

This work is supported by the National Natural Science Foundation of China under Grant Nos. 12675080, 12542053, and 12275238, the National Key Research and Development Program under Grant No. 2020YFC2201503, and the Zhejiang Provincial Natural Science Foundation of China under Grant Nos. LR21A050001 and LY20A050002, and the Fundamental Research Funds for the Provincial Universities of Zhejiang in China under Grant No. RF-A2019015. 
\bibliographystyle{apsrev4-1}
\bibliography{ref}
\end{document}